\documentclass[amsmath,prl,twocolumn,aps,superscriptaddress]{revtex4}
\usepackage{amsthm,amsfonts,graphicx,verbatim}
\pdfoutput=1

\newcommand{\ve}{\varepsilon}
\newcommand{\be}{\begin{equation}}
\newcommand{\ee}{\end{equation}}
\newcommand{\bea}{\begin{eqnarray}}
\newcommand{\eea}{\end{eqnarray}}

\newcommand{\lb}{\left[}
\newcommand{\rb}{\right]}

\newcommand{\Tr}{\textrm{Tr}}

\renewcommand{\phi}{\varphi}
\renewcommand{\epsilon}{\varepsilon}

\begin{document}

\title{Ordering of magnetic impurities and tunable electronic properties of topological insulators}
\author{D. A. Abanin}
\affiliation{Department of Physics, Princeton University, Princeton, New Jersey 08544, USA}\affiliation{Princeton Center for Theoretical Science, Princeton University, Princeton, New Jersey 08544, USA}
\author{D. A. Pesin}
\affiliation{Department of Physics, University of Texas at Austin,  Austin TX 78712 USA}

\begin{abstract}
We study collective behavior of magnetic adatoms randomly distributed on the surface of a topological insulator. As a consequence of the spin-momentum locking on the surface, the RKKY-type
interactions of two adatom spins depend on the direction of the vector connecting them, thus
interactions of an ensemble of adatoms are frustrated. We show that at low temperatures the frustrated RKKY interactions give rise to two phases: an ordered ferromagnetic phase with spins pointing perpendicular to the surface, and a disordered spin-glass-like phase. The two phases are separated by a quantum phase transition driven by the magnetic exchange anisotropy.
Ferromagnetic ordering occurs via a finite-temperature phase transition. The ordered phase breaks time-reversal symmetry spontaneously, driving the surface states into a gapped state, which exhibits an anomalous quantum Hall effect and provides a realization of the parity anomaly. We find that the magnetic ordering is suppressed by potential scattering. Our work indicates that controlled deposition of magnetic impurities provides a way to modify the electronic properties of topological insulators.

\end{abstract}
\maketitle

Topological insulators in three dimensions are a class of time-reversal-invariant materials characterized by gapless
surface states with Dirac-like dispersion (for a review, see Refs.~\cite{Kane10,Moore10} and references therein). These topologically protected
states originate from the bulk band inversion induced by strong spin-orbit interactions. The effective low-energy Hamiltonian has the form
of the Rashba spin-orbit coupling,
\be\label{eq:hamiltonian}
  H_0=v\vec n\cdot \vec p\times\vec{\sigma},
  \ee
  where $\vec{\sigma}$ is the electron spin, $v$ is the Fermi velocity, and $\vec{n}$ is the normal vector to the surface, chosen to be along $z$ direction.
The Dirac nature of the surface states is expected to manifest itself in
interesting physical effects, including magnetoelectric effect~\cite{Kane10,Moore10,Essin09,Qi}, large Kerr and universal Faraday effects~\cite{Qi,Tse10}.
In addition, locking of spin and momentum on the surface~\cite{Hsieh09}, evident from Eq. (\ref{eq:hamiltonian}), gives rise to electric
charging of magnetic textures~\cite{Nomura10}, and opens up
new opportunities for spintronics applications~\cite{Mishchenko09,Nagaosa10-sns, FranzGarate}.

\begin{figure}
\begin{center}
\includegraphics[width=3in]{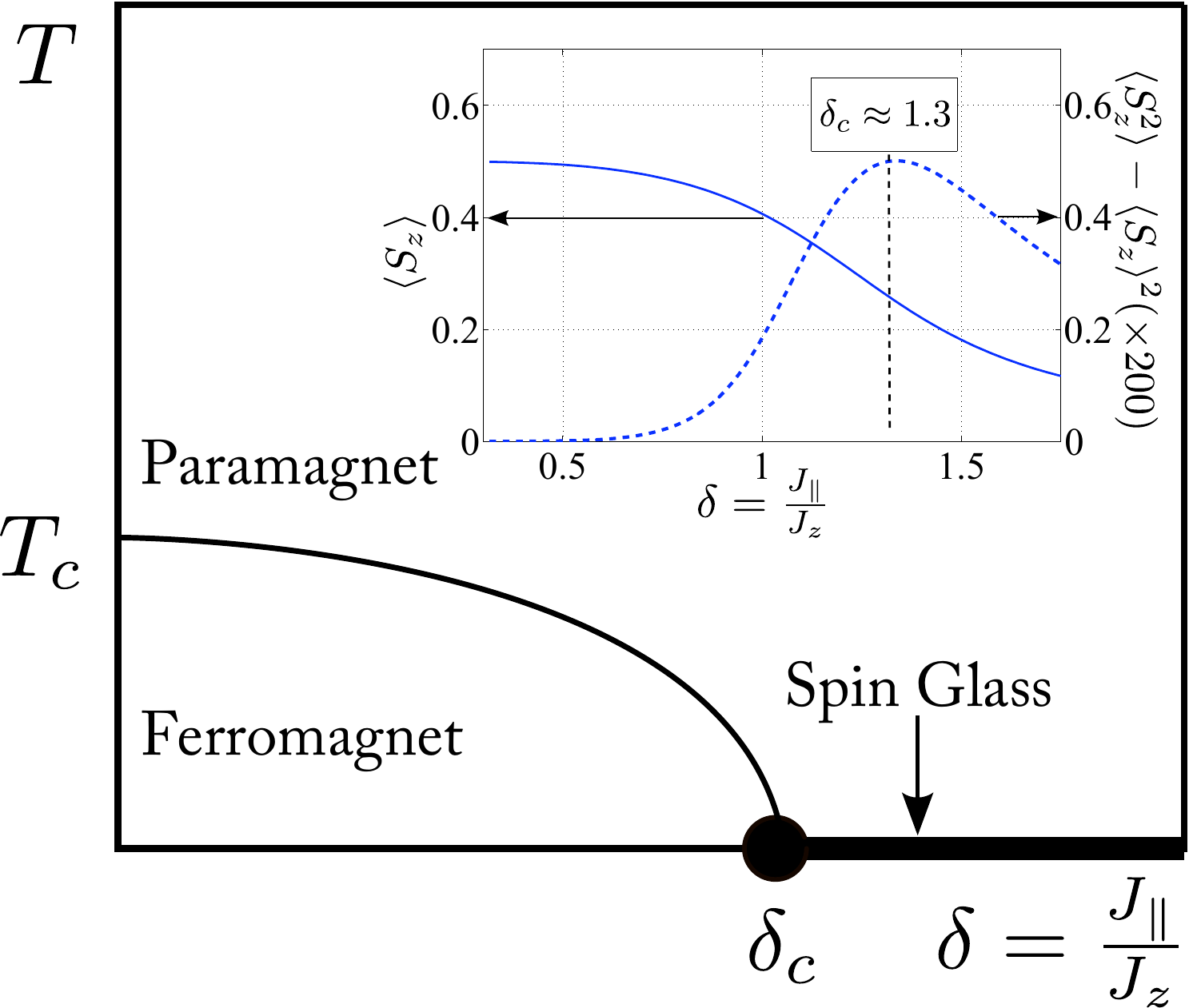}
\caption{Phase diagram of magnetic adatoms.  Inset: Magnetization of spins on the topological surface, which are interacting via RKKY interactions, as a function of the exchange anisotropy $\delta=J_{\parallel}/J_{z}$.
We find the position of the quantum critical point, $\delta_c\approx 1.3$, from the condition that the magnetization is decreased by $50 \%$. This is supported by the fluctuations of the magnetization, which exhibit a maximum at the conjectured transition point, $\delta_c \approx 1.3$. Cluster of 9 spins was considered, and averaging was performed over 150 disorder realizations. }
\label{fig:PD}
\end{center}
\end{figure}

The aforementioned physical effects and device applications rely on the ability to induce perturbations that break time reversal symmetry and open up a gap in the surface states spectrum. A relevant perturbation has the form of a mass term for Dirac electrons,
  \be\label{eq:mass}
 H_1=m\sigma_z.
 \ee
In principle, such a perturbation can be induced by depositing magnetic films. However, such a method has a significant disadvantage of
being irreversible, and likely inducing too strong magnetic fields that can completely
destroy surface states. Thus, alternative methods are needed which would allow for a controllable and reversible manipulation of the topological surfaces states.

Here we theoretically explore the possibility of  modifying electronic properties of the surface states by controlled adsorption of magnetic adatoms.
We study the collective behavior of adatom spins, determined by the RKKY-type interactions mediated by the topological surface states.  We argue that depending on the magnetic anisotropy and exchange anisotropy of a single impurity spin, the ground state of many spins is either a spin glass, or a ferromagnetically ordered state with spins pointing perpendicular to the surface. The two phases are separated by a quantum phase transition. The phase diagram of magnetic adatoms suggested in this paper is summarized in Fig.~\ref{fig:PD}. 

In the ferromagnetic phase the average exchange field of impurities induces a spin polarization dependent mass of Dirac electrons, Eq.~(\ref{eq:mass}),
opening up a band gap in the spectrum of the surface states. The mass depends on the concentration of adatoms and their type, and therefore is tunable. We find that this state is favored in a large region of the phase diagram (see below), and persists up to several tens of Kelvin. In contrast, in the spin-glass phase the average value of spin is zero, therefore, on average the time-reversal symmetry is not broken, and thus the gap is absent. There is, however, an insignificant disorder broadening which induces finite density of states at the Dirac point~\cite{unpublished}. Our work shows that
deposition of magnetic impurities provides a route to controllably change the spectral and transport properties of surface electrons.

 We start with the analysis of the RKKY interactions, employing the T-matrix description.
The interaction energy of two impurity spins located at $\vec R_{1,2}$, respectively, is given by~\cite{Shytov09}
\begin{equation}\label{eq:Omega12}
  \Omega_{12}=-T\sum_{\epsilon}\Tr\ln\lb 1-\hat t_2(\epsilon)G_0(\epsilon, \vec R)\hat t_1(\epsilon)G_0(\epsilon, -\vec R)\rb.
\end{equation}
In this expression $\vec R=\vec R_2-\vec R_1$, $\ve$ are fermionic Matsubara frequencies, $\hat t_{1,2}(\ve)$ are the low-energy T-matrices of the impurities,
and $G_0(\ve,0)$ is the unperturbed Matsubara Green's function of the surface electrons.  The trace here is taken over the spin space.
Note that there are also single-spin terms in the thermodynamic potential, which can provide easy or hard axis anisotropy for each spin, if one deals with spins larger than $1/2$. However, such anisotropy can be shown to be determined by high energies, $\ve\sim W$, and thus cannot be reliably calculated in the present approach, designed to capture low energy physics of surface electrons.

We consider a model, in which an individual impurity spin interacts with the surface electrons via anisotropic exchange Hamiltonian~\cite{Liu09},
 \be\label{eq:anisotropic_exchange}
 H_{ex}=J_z S_z \sigma_z \delta({\vec r-\vec r_0})+J_{||}\left( S_x\sigma_x+S_y\sigma_y \right) \delta({\vec r-\vec r_0}),
 \ee
where ${\vec r_0}$ is the position of the impurity, $z$ is the direction perpendicular to the surface, and $S_i$ is the impurity spin operator.
Kinetic energy of electrons is described by the Hamiltonian~(\ref{eq:hamiltonian}), with bandwidth $W$ (for Bi$_2$Se$_3$ given by $0.3\, {\rm eV}$~\cite{Kane10}), and a short range cut-off $a=v/W$.

The T-matrix, $\hat t(\ve)$ is found using the Lippmann-Schwinger equation:
\begin{equation}
  \hat t(\ve)=\vec V_i\vec\sigma+\vec V_i\vec\sigma G_0(\ve,0)\hat t(\ve),
\end{equation}
\be\label{eq:V}
 \vec V_i=\left(J_{\parallel}S^x_i,J_{\parallel}S^y_i,J_z S^z_i \right).
\ee
The expression for the unperturbed Matsubara Green's function of the surface electrons, $G_0(\ve,\vec r)$, for the Hamiltonian (\ref{eq:hamiltonian}) reads
\begin{eqnarray}\label{eq:GF}
  G_0(\ve,\vec r)=-\frac{i\epsilon}{2\pi v^2}K_0\left(\frac{|\epsilon|r}{v}\right)-\frac{i|\epsilon|}{2\pi v^2}K_1\left(\frac{|\epsilon|r}{v}\right)(\hat{r} \times \vec \sigma)_z,
\end{eqnarray}
where $K_{0,1}(x)$ are modified Bessel functions, and $\hat{r}$ is the unit vector in the direction of $\vec r$. For
$\vec{r}\to 0$ the above equation takes the following form,
\begin{equation}\label{eq:gf_coincpts}
G_0(\ve,\vec r\to 0)\equiv g(\ve)\sigma^0,  \,\,\,  g(\ve)=-\frac{i\ve}{2\pi v^2}\ln{\frac{W}{|\ve|}}.
\end{equation}
Using the above relations, we get the form of the T-matrix,
\begin{eqnarray}\label{eq:t}
  \hat t&=&t^0\sigma^0+\vec t \vec\sigma,\nonumber\\
  t^0=\frac{g\vec V^2}{1-g^2\vec V^2},\nonumber
   && \vec t=\frac{\vec V}{1-g^2\vec V^2}.
\end{eqnarray}

\begin{figure}
\begin{center}
\includegraphics[width=3.4in]{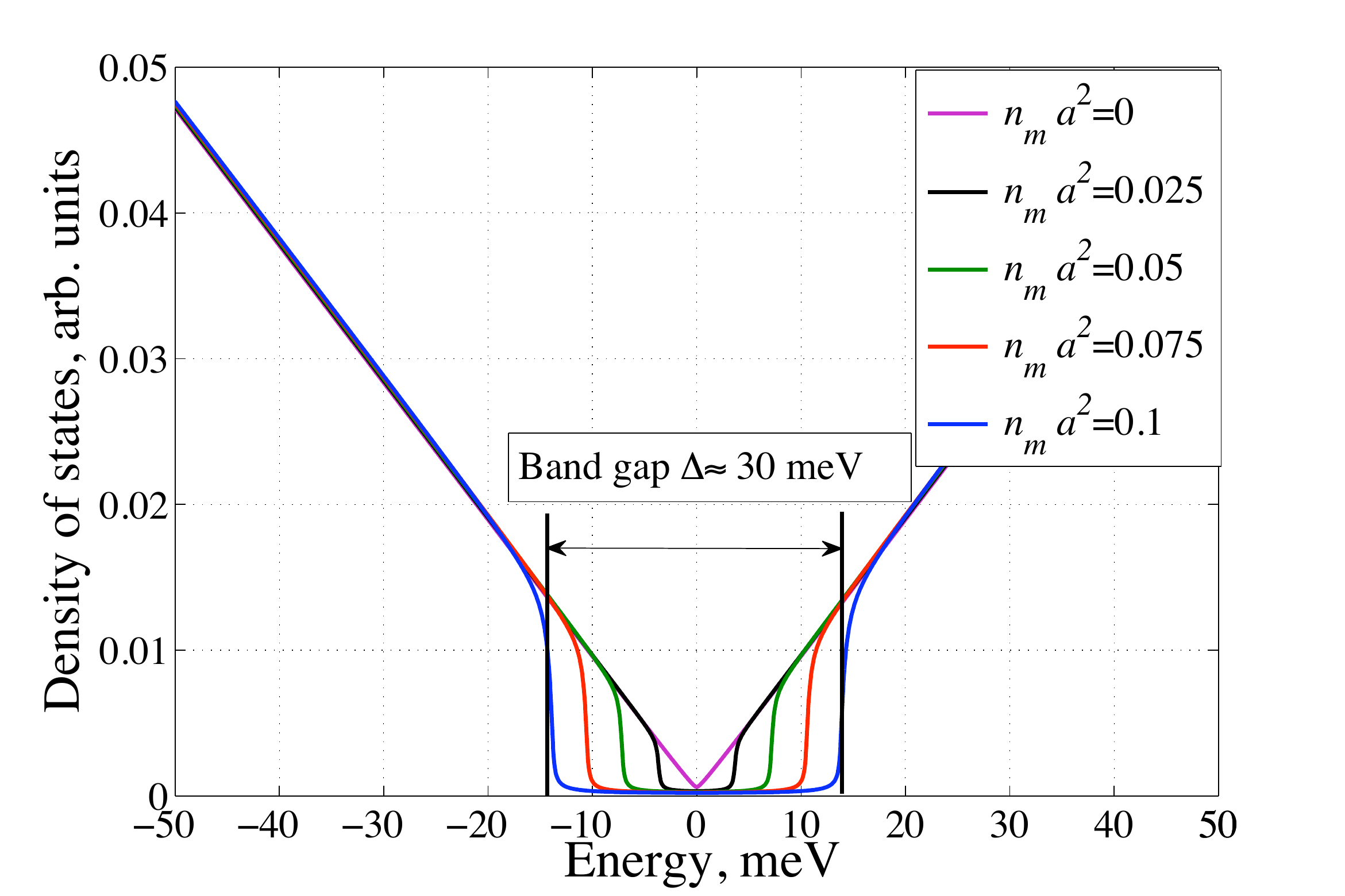}
\caption{Density of states of the topological surface states with a band gap induced by ferromagnetic ordering of adatoms. Density of states for different concentrations of magnetic adatoms. Complete polarization was assumed, and $J_zS=0.5Wa^2$, $W=0,3\, {\rm eV}$.}
\label{fig:DoS}
\end{center}
\end{figure}

The above form of the T-matrix exhibits poles at energies found from the equation $1-g^2 \vec V^2=0$. In the limit of a very large bare potential, $|\vec V|\gg W a^2$, the resonances
are positioned at low energies, $|\epsilon| \approx 2\pi v^2 /|\vec V| \ln{\frac{W |\vec {V}|}{2 \pi v^2}}$~\cite{Biswas2009}, similar to the case of graphene (see, e.g., Ref.~\cite{Shytov09}).
It is easy to see that the impurity spin dependent part of the T-matrix vanishes in both limits of $|\vec V|\to 0$ and $ |\vec V|\to\infty$.
Thus the RKKY interaction reaches a maximum at $r\sim  |\vec V|/W a\gg a$ for strong exchange, $\max (J_\parallel,J_{z})\gg 2\pi v^2/W$.  This effect is missing in the perturbative treatment of RKKY.


Lowest order perturbation theory~\cite{ketaicy} can be used for moderate exchange values, $\max (J_\parallel,J_{z})\lesssim 2\pi v^2/W$.  For simplicity, we focus on this case below; RKKY interactions for the case of strong exchange will be discussed elsewhere~\cite{unpublished}. At moderate exchange, the RKKY interactions at not too small adatom separation $r\gg a$ can be  obtained from the general expression (\ref{eq:Omega12}) by expanding the logarithm to the lowest order.
This gives, up to a small corrections of the order $a/r$,
\be\label{eq:RKKY_xyz}
U_{12}({\vec r})=-J_{z}^2\frac{C}{r^3}S_1^z S_2^z-J_{||}^2\frac{C}{r^3}({\vec S_1}\cdot{\vec r}) ({\vec S_2}\cdot{\vec r})+J_{||}^2\frac{D}{r^3} {S_1^{\perp}} S_2^{\perp},
\ee
where ${S_{1(2)}^{\perp}}={\vec S_{1(2)}}\cdot({\vec r}\times {\vec n})$, $C=\frac{1}{16\pi^3 v}\int d\xi \xi^2 (K_0^2(\xi)+K_1^2(\xi))=\frac{1}{64\pi v}$, $D=\frac{1}{16\pi^3 v}\int d\xi \xi^2 (K_1^2(\xi)-K_0^2(\xi))=\frac{1}{128\pi v}=C/2$. Therefore, the interactions between two impurities, mediated by the surface states, have an unusual, strongly anisotropic form which stems from the spin-momentum entanglement on the surface.

We analyze the collective behavior of adatoms under the realistic assumption that their spatial distribution is completely random, and the exchange coupling to the surface electrons is not too strong, as discussed above. The positional randomness combined with the form~(\ref{eq:RKKY_xyz}) of the RKKY interactions makes the in-plane interactions frustrated: the exchange is ferromagnetic between components of spins in the plane perpendicular to ${\bf r}$, and antiferromagnetic for the components of the spins parallel to ${\bf r}$.  Instead, the ferromagnetic interactions between $z$-components of spins can be optimized simultaneously. We conclude that for $\delta\equiv J_\parallel/J_z \leq 1$ the ground state of any system of adatoms is a ferromagnet with magnetization along $z$-axis. In the opposite limit, $\delta\gg 1$, the frustrated $xy$ interactions dominate,  giving rise to the ground state in which spins are frozen in the $xy$ plane, such that the average polarization in the $z$ direction vanishes. We expect this phase to be a spin glass, which is separated from an Ising-type out-of-plane ferromagnet by a quantum critical point. We also expect the ferromagnetic ground state to survive some degree of exchange anisotropy, such that the critical point corresponds to $\delta_c >1$.

As the ferromagnetic ordering breaks discrete $Z_2$ symmetry, it is separated from a paramagnetic phase by a finite-temperature second order phase transition of the conventional Ising type. The spin glass phase should not exist at finite temperature due to the fact that the surface is two dimensional. The phase diagram is summarized in Fig.~\ref{fig:PD}.

The value of $\delta_c$ depends on the magnitude of the impurity spin. We choose the most unfavorable for ferromagnetic ordering case of $S=1/2$, having the largest quantum fluctuations favoring disordered phases. To estimate the value of $\delta_c$, we have performed numerical simulations on small spin systems. We exactly diagonalized Hamiltonians of randomly distributed spin-1/2 clusters with pairwise interactions given by Eq.(\ref{eq:RKKY_xyz}). The resulting magnetization, averaged over $150$ disorder realizations, is illustrated in the inset in Fig.~\ref{fig:PD}.  At $\delta \approx 1.3$ magnetization decreases by  $50 \%$; we take this point to be the finite-size approximation to the point of quantum phase transition~\cite{Gingras08}.
The position of the transition point depends weakly on the number of spins in the cluster.
For higher impurity spins, we expect $\delta_c >1.3$; we leave the determination of the precise value $\delta_c (S)$ for future work~\cite{unpublished}.


What is the nature of the disordered phase realized at zero $T$ and $\delta\gtrsim \delta_c$? We expect that at $\delta\to \infty$ (the only interactions are in-plane) the frustrated random interaction should lead to a spin glass phase~\cite{Young86}, similarly to the case of 2D bimodal XY model~\cite{Gingras08}, and 3D dipolar dilute magnets~\cite{Dipolar09}.
We conjecture that weak ferromagnetic interaction of $z$ spin components does not destabilize the spin glass phase, and it extends to the value of $\delta=\delta_c$, where a quantum phase transition takes place. This hypothesis is supported by qualitative similarity of our system to the Sherrington-Kirkpatrick model~\cite{Sherrington}, which describes the competition between the non-frustrated ferromagnetic exchange and the frustrated sign-changing interactions. In the model of Ref.~\cite{Sherrington} ferromagnetic phase is destroyed once the frustrated part of the interactions becomes strong enough. Our model almost certainly exhibits a similar behavior. We are conducting numerical work to confirm this hypothesis~\cite{unpublished}.

Now we estimate the ferromagnetic ordering temperature. At $\delta \ll 1$, the ordering transition is essentially that of Ising spins randomly distributed in the plane and interacting via $1/r^3$ interactions. The ordering temperature can be estimated as the typical exchange interaction between two neighboring spins. 
The result for the transition temperature can be read off Eq.~(\ref{eq:RKKY_xyz}) by setting $S^{x,y}_{1,2}=0$ and $r=\sqrt{n_{m}}$, where $n_{m}$ is the concentration of impurities. This gives
an estimate
\begin{equation}\label{eq:Tc}
  T_c(n_{m})=\alpha\frac{J_z^2}{v}n^{3/2}_{m},
\end{equation}
where $\alpha$ is a numerical coefficient. Our Metropolis Monte Carlo simulations, which take into account long-range nature of interactions, confirm this estimate with $\alpha\approx 0.01$. We expect this estimate to hold for $\delta$ not too close to $\delta_c$, and, in particular, for weak exchange anisotropy, $\delta\sim 1$. For $n_ma^2\sim 1$, $J_z\sim Wa^2$, and $W=0.3 \, {\rm eV}$ (band gap of Bi$_2$Se$_3$), we obtain $T_c\sim 30\, {\rm K}$, which is in the experimentally observable range.

We now discuss the effects of the magnetic state of adatoms on the spectral and transport properties of the Dirac fermions.
The ferromagnetic phase is characterized by the spontaneous breaking of the time-reversal symmetry, and therefore leads to
a gap in the spectrum of the surface states. The most dramatic signature of the ferromagnetic ordering, detectable in transport, is that
the behavior near the neutrality point changes from metallic to strongly insulating. In contrast, both paramagnetic and spin-glass phases correspond to a gapless state of the Dirac fermions; the randomness of the potential
created by adatoms manifests itself in the small smearing of the average DOS near the Dirac point~\cite{unpublished}, similar to the case of graphene~\cite{Ostrovsky06}. The transport in this case remains metallic.

At the mean-field level, the mass induced by ordering of adatoms is given by
\be\label{eq:mass_FM}
m=n_mJ_zS.
\ee
To explore the effect of positional randomness on the gap, we have calculated density of states (DOS) in the self-consistent T-matrix approximation~\cite{AltlandSimons}, finding that
for the physically relevant case of not too strong exchange, $J_z \lesssim 2\pi v^2/W$, the mean-field result is accurate. The DOS obtained using SCTMA is displayed in Fig.~\ref{fig:DoS}. Taking
$J_z /a^2 \approx 300\, {\rm meV}$, $na^2\approx 0.1$, and $S=1/2$, we obtain the estimate of mass $m\approx 15\, {\rm meV}$, which puts it in the experimentally observable range.


An important consequence of a mass in the spectrum Dirac fermions is quantized half-integer Hall conductivity~\cite{Semenoff83},
 even in the absence of magnetic field. Such anomalous quantum Hall effect (QHE) is a direct consequence
of the parity anomaly and has been predicted a while ago~\cite{Jackiw84}, but could not be observed experimentally to date. The magnetic ordering on the surface provides a way to experimentally observe anomalous QHE. This can be done, e.g., in a thin slab geometry, where spins polarize in the same direction on the opposite sides of the slab. In such a geometry, half-integer Hall conductivities of the two surface add, giving a quantized Hall conductivity $e^2/h$, which is observable in the standard Hall-bar geometry. An alternative way to realize anomalous QHE in topological insulators, proposed recently on the basis of the ab initio calculations, involves
changing the bulk band structure with magnetic atoms~\cite{Zhang10}.

Now we analyze the effect of potential impurities on the interactions and ordering of magnetic impurities -- the issue relevant for current experiments, where potential disorder is inevitably
introduced by chemical doping. It is well known in the usual case of disordered metals that the impurity averaged exponentially decaying RKKY interaction does not describe interaction between magnetic impurities~\cite{Zyuzin}. Instead, the sample-specific interaction in a disordered sample has a powerlaw behavior.

In our case, disorder also causes exponential decay of the averaged RKKY interactions at large distances.
The relevant length scale is given by (twice) the Green's function decay length at $\ve\to 0$. In the case of strong nonmagnetic impurities, this length scale is given by~\cite{Shytov09,Ostrovsky06}
\be
\ell_{dis}\sim\frac{1}{\sqrt{n_p}}\log\frac{W^2}{v^2n_p},
\ee
where $n_p$ is the density of nonmagnetic impurities, and we neglected factors of order of unity. We find~\cite{unpublished} that RKKY interactions decays exponentially at distances $r\gg \ell_{dis}$. The sample-specific RKKY interactions are random, have the Dzyaloshinskii-Moriya form, and decay as a powerlaw~\cite{Zyuzin}.  We thus are led to the conclusion that for large densities of nonmagnetic impurities, $n_p\gg n_m$, the RKKY interaction is primarily of the Dzyaloshinskii-Moriya form, and the impurity spins remain magnetically disordered down to very low temperatures and/or small values of $J_\parallel/J_z$. Thus, magnetic ordering can only be observed in relatively clean samples.

Finally, we discuss the effect of finite doping on the magnetic ordering of adatoms.
In the doped system, the phase diagram discussed above remains essentially unchanged for large enough adatom concentration, such that adatom separation is smaller than the Fermi wave length,
$n_m^{-1/2}\ll \lambda_F$.
At small adatom concentration $n_m^{-1/2}\gtrsim \lambda_F$, the Fermi surface effects make RKKY coupling sign-changing.
In addition, as pointed out in Ref.~\cite{ketaicy}, in this regime the RKKY interaction acquires a Dzyaloshinskii-Moriya component, which frustrates interactions even more.
Thus a finite
doping stabilizes the spin glass phase and suppresses the ferromagnetic ordering.


In conclusion, we have studied collective properties of magnetic impurities on a topological surface. The ferromagnetic ordering, occurring in particular for the isotropic exchange, provides a way to
gap out the surface state and realize an anomalous Quantum Hall state.
This requires clean samples, with concentration of potential impurities being smaller than the concentration of magnetic adatoms.

Recently, we have become aware of an experimental work~\cite{Wray10}, where deposition of magnetic impurities was used as a tool to modify properties of Bi$_2$Se$_3$. Above certain density, ferromagnetic ordering, leading to a gap opening, was observed, in agreement with our analysis above. The suppression of ordering at low densities $n_m$ observed experimentally is likely due to potential disorder, which stabilizes the spin glass phase, as discussed above.


\textbf{Acknowledgements.} We thank L. Balents, L. Glazman, M. Gingras, D. Huse, A. H. MacDonald, S. Sondhi, and B. Z. Spivak for illuminating discussions, and L. A. Wray and M. Z. Hasan for very helpful discussion of experiment. DAA thanks Aspen Center for Physics, where part of this work was completed, for hospitality. DAP's work was supported by Welch Foundation grant F1473, and by the ARO MURI on bioassembled nanoparticle arrays. 


\begin{references}
\bibitem{Kane10}
M. Z. Hasan, C. L. Kane,  arXiv:1002.3895.

\bibitem{Moore10}
J. Moore, Nature 464, 194-198 (2010).

\bibitem{Essin09}
A. M. Essin, J. E. Moore, D. Vanderbilt, Phys. Rev. Lett. 102, 146805 (2009).





\bibitem{Qi}
X.-L. Qi, T. L. Hughes, S.-C. Zhang, Phys. Rev. B \textbf{78}, 195424 (2008)


\bibitem{Tse10}
W.-K. Tse, A. H. MacDonald, Phys. Rev. Lett. 105, 057401 (2010).

\bibitem{Hsieh09} D. Hsieh et.al., Nature 452, 970 (2008);  D. Hsieh et al., Nature 460, 1101-1105 (2009).

\bibitem{Nomura10}
K. Nomura, N. Nagaosa, arXiv:1006.4217.


\bibitem{Mishchenko09}
P.G. Silvestrov, E.G. Mishchenko, arXiv:0912.4658 (2009).


\bibitem{Nagaosa10-sns}
Y. Tanaka {\it et al.}, Phys. Rev. Lett. 103, 107002 (2009);  T. Yokoyama, Y. Tanaka, N. Nagaosa, Phys. Rev. Lett. 102, 166801 (2009).

\bibitem{FranzGarate}
I. Garate, M. Franz, Phys. Rev. Lett. 104, 146802 (2010).

\bibitem{Liu09}
Q. Liu {\it et al.}, Phys. Rev. Lett. 102, 156603 (2009).


\bibitem{Shytov09}
A. V. Shytov, D. A. Abanin, L. S. Levitov, Phys. Rev. Lett. {\bf103}, 016806 (2009).



\bibitem{Sherrington}
D. Sherrington and S. Kirkpatrick, Phys. Rev. Lett. {\bf 35}, 1792 (1975).

\bibitem{Young86}
K. Binder and A. P. Young
Rev. Mod. Phys. {\bf 58}, 801 (1986).

\bibitem{unpublished}
D. A. Pesin and D. A. Abanin, in preparation.

\bibitem{Ostrovsky06}
P. M. Ostrovsky, I. V. Gornyi, A. D. Mirlin,
Phys. Rev. B {\bf74}, 235443 (2006).

\bibitem{Gingras08}
M. Weigel, M. J. P. Gingras, Phys. Rev. B 77, 104437 (2008).

\bibitem{Dipolar09}
P. Stasiak, M. J. P. Gingras,
arXiv:0912.3469 (2009).

\bibitem{Biswas2009}
R. Biswas, A. Balatsky,  arXiv:0910.4604.


\bibitem{ketaicy}
F. Ye {\it et al.}, Europhys. Lett. 90, 47001 (2010).



\bibitem{Semenoff83}
A. J. Niemi and G. W. Semenoff, Phys. Rev. Lett. 51, 2077 (1983).




\bibitem{Jackiw84}
R. Jackiw, Phys. Rev. D {\bf29}, 2375 (1984).



\bibitem{Zhang10}
R. Yu {\it et al.}, Science 329, 61 (2010).

\bibitem{Zyuzin}
A. Y. Zyuzin and B. Z. Spivak, JETP Lett., {\bf 43}, 234 (1986); L. N. Bulaevskii  and S. V. Panyukov, JETP Lett., {\bf 43}, 240 (1986); G. Bergmann, Phys. Rev. B, {\bf36}, 2469 (1987).




\bibitem{AltlandSimons}
A. Altland and B. Simons, \textit{Condensed Matter Field Theory} (Cambridge University Press, 2006)

\bibitem{Wray10}
L.A. Wray et.al., Nature Phys. (in press) (2010).

\end{references}
\end{document}